\documentclass{article}
\usepackage[a4paper, total={14.5cm, 24.5cm}]{geometry}

\usepackage{comment}
\usepackage{amsmath}
\usepackage{hyperref}
\usepackage{cleveref}
\usepackage{mwe}
\usepackage{todonotes}
\usepackage{siunitx}
\usepackage{listings}
\usepackage{xcolor}

\usepackage[backend=bibtex]{biblatex}
\definecolor{codegreen}{rgb}{0,0.6,0}
\definecolor{codegray}{rgb}{0.5,0.5,0.5}
\definecolor{codepurple}{rgb}{0.58,0,0.82}
\definecolor{backcolour}{rgb}{0.95,0.95,0.92}
\lstdefinestyle{mystyle}{
    backgroundcolor=\color{backcolour},
    commentstyle=\color{codegreen},
    keywordstyle=\color{magenta},
    numberstyle=\tiny\color{codegray},
    stringstyle=\color{codepurple},
    basicstyle=\ttfamily\footnotesize,
    breakatwhitespace=false,
    breaklines=true,
    captionpos=b,
    keepspaces=true,
    numbers=left,
    numbersep=5pt,
    showspaces=false,
    showstringspaces=false,
    showtabs=false,
    tabsize=2
}
\title{FEniCSx-pctools: Tools for PETSc block linear algebra preconditioning in FEniCSx}
\author{Martin \v{R}eho\v{r}\thanks{\url{martin.rehor@rafinex.com} Rafinex S.\`{a} r.l., Luxembourg.} \and Jack S.
Hale\thanks{\url{jack.hale@uni.lu} Institute of Computational Engineering, Department of Engineering, Faculty of Science, Technology and Medicine,
University of Luxembourg, Luxembourg.}}

\begin{document}
\maketitle

\section*{Abstract}
Solving partial differential equations with the finite element method leads to
large linear systems of equations that must be solved. When these systems have
a natural block structure due to multiple field variables, using iterative
solvers with carefully designed preconditioning strategies that exploit the
underlying physical structure becomes necessary for an efficient and scalable
solution process. FEniCSx Preconditioning Tools (FEniCSx-pctools) is a software
package that eases the specification of PETSc (Portable, Extensible Toolkit for
Scientific Computation) block preconditioning strategies on linear systems
assembled using the DOLFINx finite element solver of the FEniCS Project. The
package automatically attaches all necessary metadata so that preconditioning
strategies can be applied via PETSc's standard options database to monolithic
and block assembled systems. The documented examples
include a simple mixed Poisson system and more complex pressure
convection-diffusion approach to preconditioning the Navier-Stokes equations.
We show weak parallel scaling on a fully coupled temperature-Navier-Stokes
system up to \num{8192} MPI (Message Passing Interface) processes,
demonstrating the applicability of the approach to large-scale problems.

FEniCSx-pctools is available under the LGPLv3 or later license and is developed
on GitLab \url{https://gitlab.com/rafinex-external-rifle/fenicsx-pctools}. The
documentation is available at
\url{https://rafinex-external-rifle.gitlab.io/fenicsx-pctools/}.

\section*{Keywords}
finite elements, iterative solvers, preconditioning, mixed systems,
multiphysics simulations.

\section*{Introduction}

Solving linear systems is a core step in the numerical solution of partial
differential equations (PDE). To solve large linear systems efficiently on
high-performance computers, iterative solvers are essential. However, iterative
solvers require effective preconditioning to be viable; without this,
convergence may be unacceptably slow, or fail~\cite{allaire_numerical_2008}.
When modelling complex physical phenomena involving multiple coupled fields,
such as incompressible fluid flow, mixed formulations of elasticity, and
electromagnetism, the linear systems have a block structure where the different
variables have natural couplings. It is now well established that the iterative
solution of block linear systems requires block preconditioning strategies that
exploit the structure of the coupled
problems~\cite{benzi_numerical_2005,elman_finite_2014}.

The purpose of this software metadata paper is to describe FEniCSx Preconditioning
Tools (henceforth FEniCSx-pctools), an add-on for
DOLFINx~\cite{baratta_dolfinx_2023} and PETSc (Portable, Extensible Toolkit for
Scientific Computation)~\cite{petsc-user-ref} that eases the specification of
complex PETSc-based block-structured preconditioners and linear solvers.
FEniCSx-pctools requires both DOLFINx and PETSc/petsc4py and was produced as a
supporting tool in a larger research project focused on the topology optimisation
of fluidic devices.

The main contribution of FEniCSx-pctools is to bridge the gap between DOLFINx
and PETSc by offering a set of algorithms that can analyse the high-level
Unified Form Language (UFL)~\cite{alnaes_unified_2014} representation of a
block-structured finite element problem and subsequently attach the necessary
PETSc metadata to the existing DOLFINx assembled monolithic and block assembled linear
algebra objects. With this metadata, advanced block preconditioning strategies
can be specified straightforwardly using PETSc's standard options-based
configuration system~\cite{brown_composable_2012}.

We include three fully documented demos. The first shows a Schur complement
preconditioner for the mixed Poisson problem, based on a design proposed
in~\cite{rusten_interior_1996}. The second sets up a Schur complement
preconditioner of the velocity-pressure Navier-Stokes equations using the
pressure-convection-diffusion (PCD) approach proposed
in~\cite{elman_finite_2014}. In the final demo, an elementary system of
algebraic equations is used to illustrate the ability to change the solver
configuration at runtime independently of the model formulation.

\section*{Related work}\label{sec:related}
We focus on software that provides a high-level interface to block
preconditioning strategies in lower-level sparse linear algebra libraries like
Trilinos~\cite{trilinos-website,baker_tpetra_2012} and PETSc~\cite{petsc-user-ref}.

\begin{enumerate}
\item The Firedrake Project~\cite{rathgeber_firedrake_2016} builds directly on
top of the PETSc DM package~\cite{lange_efficient_2016} allowing for
the straightforward specification of block-structured algebraic
systems and composable physics-based preconditioners.

\item CBC.Block~\cite{mardal_block_2012} provides block preconditioning tools
within the legacy DOLFIN library~\cite{logg_dolfin_2010,alnaes_fenics_2015} using the Trilinos
linear algebra backend. A particularly strong aspect of CBC.Block is
its domain-specific algebraic language for specifying block linear
algebra preconditioners. CBC.Block is one of the core components of
HAZniCS~\cite{budisa_haznics_2023}, a software toolbox for solving
interface-coupled multiphysics problems.

\item FENaPack~\cite{fenapackpaper} is a preconditioning package for the legacy
DOLFIN library using the PETSc linear algebra backend. A particular
focus is on implementing the PCD approach for preconditioning the
Navier-Stokes equations.

\item PFIBS~\cite{chang_pfibs_2022}, like FENaPack, is a parallel
preconditioning package for the legacy DOLFIN library using the
PETSc linear algebra backend. It contains a class BlockProblem that can
split a monolithic PETSc matrix into blocks. PFIBS also contains an
example of PCD preconditioning of the Navier-Stokes equations.
\end{enumerate}

The first-mentioned finite element solver,
Firedrake~\cite{rathgeber_firedrake_2016}, offers block preconditioning as
standard. The remaining packages are extensions to the legacy DOLFIN finite
element solver that ease block preconditioning in Trilinos or PETSc.
FEniCSx-pctools is similar in scope to these remaining packages, in that it
extends the basic capabilities of the new DOLFINx finite element
solver~\cite{baratta_dolfinx_2023}, which is linear algebra backend agnostic
and does not use PETSc's DM functionality like Firedrake, to support
PETSc-based block preconditioning. Additionally, like FENaPack and PFIBS,
FEniCSx-pctools contains an implementation of PCD preconditioning which is
non-trivial to implement for users.

\section*{Implementation and architecture}
In this section we briefly outline the structure of FEniCSx-pctools library
before illustrating its use via the example of preconditioning a mixed Poisson
problem. We assume basic knowledge of the mathematics of the finite element
method, the DOLFINx finite element solver and PETSc, see
e.g.~\cite{fenicsx_tutorial_2025}.

Before continuing we briefly explain the three main types of matrix assembly
routines that DOLFINx offers in the current development version
\texttt{0.10.0.dev0} and whether they allow for block preconditioning.

The first type, monolithic, is created when assembling via
\texttt{assemble\_matrix(..., kind=}
\texttt{None)} a finite element form defined on a
single \texttt{dolfinx.fem.FunctionSpace} containing a single
\texttt{basix.ufl.FiniteElement} or a single \texttt{basix.ufl.MixedElement}
containing multiple \texttt{basix.ufl.FiniteElement} objects (one for each
variable). In the latter case, the degree-of-freedom maps related to each
variable are interleaved into a single global degree-of-freedom map, resulting
in strong data locality, but the assembled matrix loses the `natural' block
structure and block preconditioning is not available.

The second type, nest, is created when assembling a
finite element form defined on a sequence of
\texttt{dolfinx.fem.FunctionSpace} objects using the
\texttt{assemble\_matrix(..., kind="nest")} call. Each
interaction between a pair of variables (including
self-interactions) is assembled into its own matrix and the
global system is represented as a PETSc MATNEST matrix composed of
these submatrices. The advantage of the nest type is that
block preconditioning is supported natively, but in the case
the user wants to use a direct solver, only MUMPS can be used.

The third type, block, is created when assembling a finite element
form defined on a sequence of \texttt{dolfinx.fem.FunctionSpace} objects using
the \texttt{assemble\_matrix(..., kind="mpi")} call. In this case, the
degree-of-freedom maps for each variable occupy a contiguous block in a global
degree-of-freedom map. The forms are assembled into a standard PETSc MATMPIAIJ matrix and
the block structure is retained. An advantage of the block type is that any
direct solver can be used for solution, not just MUMPS, as in nest. However,
as the degree-of-freedom map has lost explicit information about the variables,
block preconditioning is not available.

The key contribution of FEniCSx-pctools is to enable block preconditioning for
monolithic and block assembly.

FEniCSx-pctools is implemented as a Python package \texttt{fenicsx-pctools}
with two subpackages named \texttt{mat} (matrix, abbreviated as in PETSc to mat)
and \texttt{pc} (preconditioners, abbreviated as in PETSc to pc). The
package is fully documented and is type-hinted/checked.

The \texttt{mat} subpackage primarily contains two fundamental factory functions named
\texttt{create\_}
\texttt{splittable\_matrix\_block} and \texttt{create\_splittable\_matrix\_monolithic},
which accept a\linebreak DOLFINx assembled PETSc \texttt{Mat}, of either block or monolithic type,
plus the associated UFL form and return a new PETSc Python-context \texttt{Mat}
that allows for the matrix to be split into submatrices (blocks) corresponding to the original
field variables.

The \texttt{pc} subpackage primarily contains a class \texttt{WrappedPC}, a PETSc
Python-context \texttt{PC} that allows for preconditioners to work with the
PETSc Python-context \texttt{Mat} created by the factory functions in the
aforementioned \texttt{mat} subpackage. The class \texttt{WrappedPC} is not used
directly; instead, users tell PETSc to use it as a Python-context preconditioner and combine it
with a standard PETSc \texttt{PC}. More specifically, the Python-context preconditioner sets the
standard preconditioner as its own attribute and uses it to interact with the underlying block
matrix. This way it can wrap any type of preconditioner (hence the generic name), but it is
designed to extend the configurability of the fieldsplit preconditioner.
Additionally, \texttt{pc} contains implementations of two PCD
preconditioning strategies \texttt{PCDPC\_vX} and \texttt{PCDPC\_vY}. All of
the prior mentioned preconditioner classes inherit from a shared base class \texttt{PCBase}
which can be used by advanced users to create their own preconditioners.

\begin{figure}[!h]
\centering
\vspace{2mm}
\includegraphics[width=\textwidth]{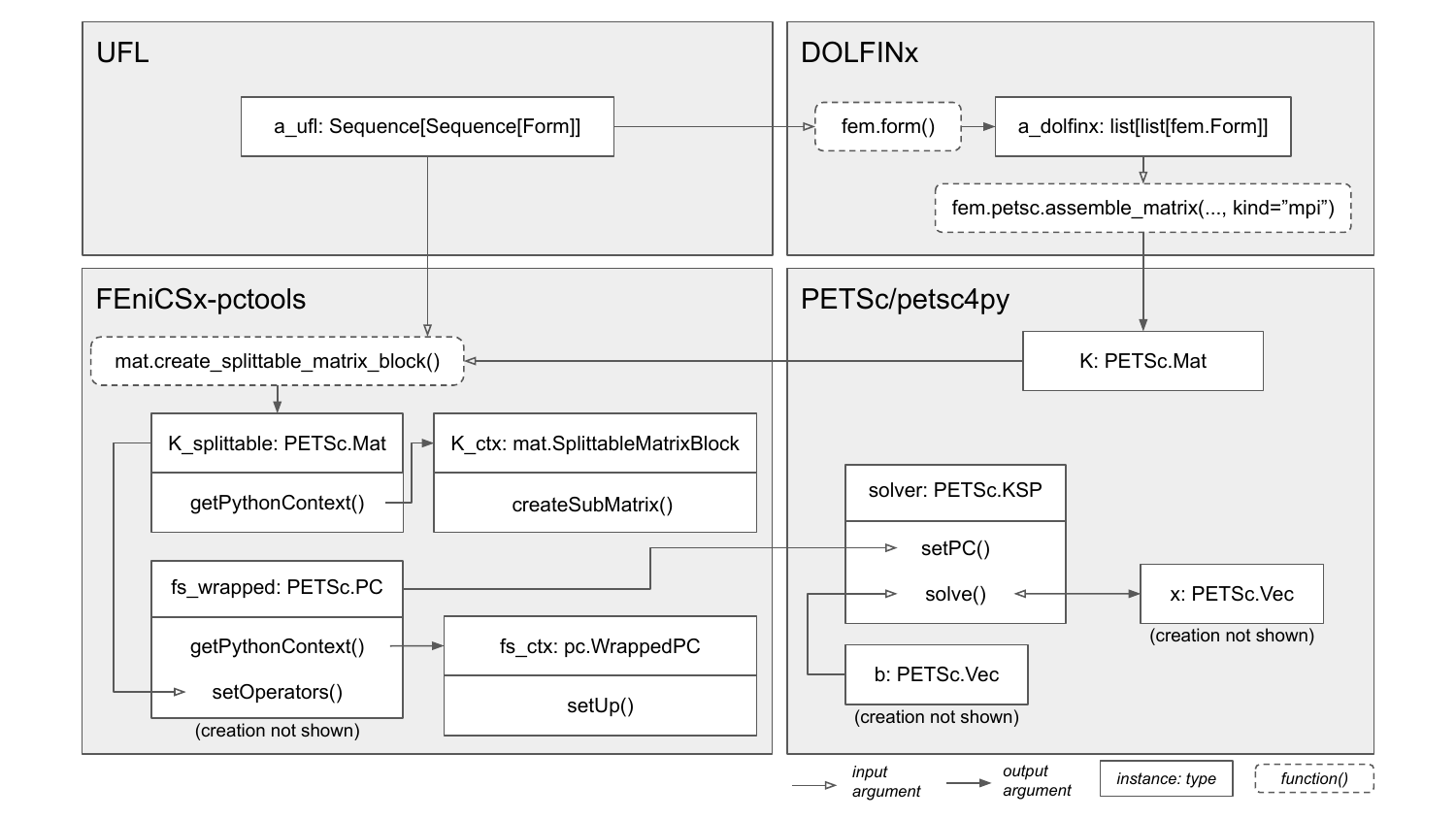}
\caption{Diagram summarising the workflow to implement a custom fieldsplit
	preconditioner using the FEniCSx-pctools library. (Top left block, UFL)
	The symbolic UFL bilinear form defining a block operator is first
	defined as a 2D list of UFL forms \texttt{a\_ufl}. (Top right block,
	DOLFINx) The list of UFL forms is just-in-time compiled to a list of
	compiled finite element forms \texttt{a\_dolfinx}, which is then
	assembled into a block matrix \texttt{K} in PETSc’s ``mpiaij''
	format (bottom right block, PETSc/petsc4py). We now enter the domain of
	FEniCSx-pctools, bottom left block. Instead of directly associating the
	matrix \texttt{K} with the linear \texttt{solver}, we first pass
	the original UFL form \texttt{a\_ufl} and \texttt{K} to
	\texttt{create\_splittable\_matrix\_block} to create a splittable matrix
	\texttt{K\_splittable}. This is a necessary step to allow block extraction based on
	arbitrarily combined index sets provided by the fieldsplit
	preconditioner \texttt{fs\_wrapped}. The extraction itself is implemented in the
	\texttt{createSubMatrix} method of the splittable matrix’
	Python-context \texttt{K\_ctx}, while the index sets are configurable
	via \texttt{setUp} method of the preconditioner’s Python-context
	\texttt{fs\_ctx}. (Bottom right block, PETSc/petsc4py) The methods in
	\texttt{fs\_ctx} and \texttt{K\_ctx} are transparently executed by the linear
	\texttt{solver} upon calling its \texttt{solve} method with the right-hand side
	vector \texttt{b} and a vector \texttt{x} to store the solution.}\label{fig:diagram}
\end{figure}

The diagram in \cref{fig:diagram} illustrates the role of FEniCSx-pctools in the FEniCS
ecosystem, and shows how to set up a fieldsplit preconditioner to allow block preconditioning
of a block matrix.

We now move onto an example for DOLFINx version \texttt{0.10.0.dev0} showing the use of
\texttt{create\_splittable\_matrix\_block} and \texttt{WrappedPC} . A classic
example of a block-structured linear system arises from the finite element
discretisation of the mixed Poisson problem. Preconditioning this problem is a
foundational example and so is widely used to show the basic aspects of
preconditioning in software frameworks. In weak form we seek a vector-valued
flux $q_h \in Q_h$ and a scalar-valued pressure $p_h \in P_h$, such that
\begin{subequations}
\begin{align}
	\left( q_h, \tilde{q} \right) + \left( p_h, \mathrm{div}(\tilde{q}) \right) &= 0 \quad &\forall \tilde{q} \in Q_h,\\
	\left( \mathrm{div}(q_h), \tilde{p} \right) &= \left( -f, \tilde{p} \right) \quad &\forall \tilde{p} \in P_h,
\end{align}
\end{subequations}
where $\left( \cdot, \cdot \right)$ denotes the usual $L^2$ inner product on
the finite element mesh and $f$ a known forcing term. In discrete block form
this can be written as a saddle point linear system with unknown vectors of
finite element coefficients $q$ and $p$
\begin{equation}\label{eq:mixed_poisson_block}
\begin{bmatrix}
A & B^T \\
B & O
\end{bmatrix}
\begin{bmatrix}
q \\ p
\end{bmatrix}
=
\begin{bmatrix}
0 \\ g
\end{bmatrix},
\end{equation}
where $A$ is a square matrix arising from the bilinear form $(q_h, \tilde{q})$,
$B$ and $B^T$ are non-square matrices arising from the bilinear forms
$(\mathrm{div}(q_h), \tilde{p})$ and $(p_h, \mathrm{div}(\tilde{q}))$
respectively, $O$ is a square matrix of zeros, $0$ is a vector of zeros and $g$
is a vector arising from the linear form $(-f, \tilde{p})$. Finally, we denote
the block matrix on the left of \cref{eq:mixed_poisson_block} $K$, the block
vector of flux and pressure unknowns $x$, and the block vector on the right $b$,
i.e. $Kx = b$.

The matrix $K$ and the vector $b$ can be assembled in DOLFINx using the standard
code shown in \cref{fig:standard_code}.
At this point DOLFINx has assembled the left-hand side matrix of
\cref{eq:mixed_poisson_block} in two steps:
\begin{enumerate}
	\item By creating a PETSc matrix based on the array of bilinear forms; this
		involves the creation of a merged sparsity pattern with the block
		layout dictated by the arrangement of forms in the array.
	\item Assembling the forms into the matrix initialised in step 1, in a
		blockwise manner; this involves the creation of index sets that can
		be used to extract and assemble each individual submatrix.
\end{enumerate}

\begin{figure}
\begin{lstlisting}[language=python,linerange=1-40,firstnumber=1]
from mpi4py import MPI
from petsc4py import PETSc

import numpy as np

from basix.ufl import element
from dolfinx import fem, mesh
from dolfinx.fem.petsc import assemble_matrix, assemble_vector
from fenicsx_pctools.mat import create_splittable_matrix_block
from ufl import CellDiameter, FacetNormal, Measure, TestFunction, TrialFunction, ZeroBaseForm, avg, div, ds, dS, grad, inner, jump

domain = mesh.create_unit_square(MPI.COMM_WORLD, 1024, 1024, mesh.CellType.quadrilateral)

k = 1
Q_el = element("BDMCF", domain.basix_cell(), k)
P_el = element("DG", domain.basix_cell(), k - 1)
Q = fem.functionspace(domain, Q_el)
P = fem.functionspace(domain, P_el)

q = TrialFunction(Q)
q_t = TestFunction(Q)

p = TrialFunction(P)
p_t = TestFunction(P)

f = fem.Function(P)
rng = np.random.default_rng()
f.x.array[:] = rng.uniform(size=f.x.array.shape)

dx = Measure("dx", domain)
a = [[inner(q, q_t) * dx, inner(p, div(q_t)) * dx], [inner(div(q), p_t) * dx, None]]
L = [ZeroBaseForm((q_t,)), -inner(f, p_t) * dx]
a_dolfinx = fem.form(a)
L_dolfinx = fem.form(L)

K = assemble_matrix(a_dolfinx, kind="mpi")
K.assemble()

b = assemble_vector(L_dolfinx, kind="mpi")
b.ghostUpdate(addv=PETSc.InsertMode.INSERT, mode=PETSc.ScatterMode.FORWARD)
\end{lstlisting}
\caption{Standard DOLFINx code for assembling the block linear system $Kx = b$
	for the mixed Poisson problem. We define a mesh consisting of $1024
	\times 1024$ quadrilateral cells. For the flux space $V_h$ we choose
	Brezzi-Douglas-Marini elements of first-order, and for the pressure
	space $Q_h$ discontinuous Lagrange elements of zeroth-order. The right-hand
	side forcing term $f$ is drawn from a uniform distribution and then the
	mixed Poisson variational problem is defined using UFL.}\label{fig:standard_code}
\end{figure}

\begin{figure}
\begin{lstlisting}[language=python,linerange=42-43,firstnumber=42]
K_splittable = create_splittable_matrix_block(K, a)
K_splittable.setOptionsPrefix("mp_")
\end{lstlisting}
\caption{Continuation of \cref{fig:standard_code}. The fundamental FEniCSx-pctools factory function
	\texttt{create\_splittable\_matrix\_block} takes the DOLFINx assembled matrix along with the
	array of bilinear forms that defines it, and returns a PETSc Mat of type ``python'' with the
	necessary functionality to apply block preconditioning strategies.}\label{fig:pctools_code}
\end{figure}

The extraction of a submatrix is the key mechanism exploited by PETSc's fieldsplit-based
preconditioners \cite{brown_composable_2012}. The splittable block matrix created with the
dedicated factory function, see \cref{fig:pctools_code}, is equipped with metadata including
row and column index sets defining the individual blocks. These can be arbitrarily combined
(concatenated) and used to define a fieldsplit group to be extracted. The grouping is not
limited to neighbouring blocks, which is the case when using PETSc MATNEST matrices directly
with the fieldsplit preconditioner.

We now describe and specify an upper-diagonal Schur complement preconditioner using PETSc.
Writing the $LDU$ decomposition of $K$ gives
\begin{equation}\label{eq:ldu}
K = LDU =
\begin{bmatrix}
I & 0 \\
BA^{-1} & I
\end{bmatrix}
\begin{bmatrix}
A & 0 \\
0 & S
\end{bmatrix}
\begin{bmatrix}
I & A^{-1}B^T \\
0 & I
\end{bmatrix}.
\end{equation}
Choosing to use the inverse of the diagonal $D$ and upper $U$ components of \cref{eq:ldu}
as a preconditioner
\begin{equation}
P_{\mathrm{upper}} = DU =
\begin{bmatrix}
A & 0 \\
0 & S
\end{bmatrix}
\begin{bmatrix}
I & A^{-1}B^T \\
0 & I
\end{bmatrix},
\end{equation}
leads to the following \emph{upper} Schur complement \emph{left}
preconditioned block system of equations
\begin{equation}\label{eq:preconditioned}
P_{\mathrm{upper}}^{-1} K x = P_{\mathrm{upper}}^{-1} b,
\end{equation}
Use of GMRES and upper Schur complement preconditioning can be specified using
the PETSc options shown in \cref{fig:petsc_wrapped_code}. Note how we use
\texttt{WrappedPC} as a high-level preconditioner which acts as an embedded layer for
reading the precise fieldsplit configuration from the options database and completing
the metadata that will be eventually used for the extraction of the submatrices.

\begin{figure}[!b]
\begin{lstlisting}[language=python,linerange=45-63,firstnumber=45]
solver = PETSc.KSP().create(MPI.COMM_WORLD)
solver.setOptionsPrefix("mp_")
solver.setOperators(K_splittable)

options = PETSc.Options()
options.prefixPush("mp_")
options["ksp_type"] = "gmres"
options["ksp_rtol"] = 1e-8
options["ksp_monitor_true_residual"] = ""
options["pc_type"] = "python"
options["pc_python_type"] = "fenicsx_pctools.pc.WrappedPC"

options.prefixPush("wrapped_")
options["pc_type"] = "fieldsplit"
options["pc_fieldsplit_type"] = "schur"
options["pc_fieldsplit_schur_fact_type"] = "upper"
options["pc_fieldsplit_schur_precondition"] = "user"
options["pc_fieldsplit_0_fields"] = "0"
options["pc_fieldsplit_1_fields"] = "1"
\end{lstlisting}
\caption{Continuation of \cref{fig:pctools_code}. We first specify that PETSc should use the
	custom class \texttt{fenicsx\_pctools.pc.WrappedPC} to wrap the fieldsplit preconditioner,
	and GMRES as an outer solver for $P^{-1}_{\mathrm{upper}}Kx = P^{-1}_{\mathrm{upper}}b$.
	At the next level, we ask for upper Schur complement preconditioning and that the structure
	of the preconditioner will be specified by the user.}\label{fig:petsc_wrapped_code}
\end{figure}

In the general case, $S$ is a dense matrix that cannot be stored explicitly, let
alone inverted. To avoid this, we suppose the existence of a `good' approximate
action for both $A^{-1} \approx \tilde{A}^{-1}$ and $S^{-1} \approx
\tilde{S}^{-1}$, i.e. we substitute
\begin{equation}\label{eq:P_inverse}
	\tilde{P}_{\mathrm{upper}}^{-1} =
\begin{bmatrix}
I & -\tilde{A}^{-1}B^T \\
0 & I
\end{bmatrix}
\begin{bmatrix}
\tilde{A}^{-1} & 0 \\
0 & \tilde{S}^{-1}
\end{bmatrix},
\end{equation}
for $P_{\mathrm{upper}}^{-1}$ in \cref{eq:preconditioned}, where the tilde
$(\tilde{\cdot})$ denotes an approximate (inexact) inverse.

To compute \cref{eq:P_inverse} we still must specify the form of both
$\tilde{A}^{-1}$ and $\tilde{S}^{-1}$. One reasonable choice is to take
$\tilde{A}^{-1}$ as a single application of a block Jacobi preconditioned
inverse mass matrix on the finite element flux space $Q_h$. Note that this mass
matrix is in fact the already assembled upper left block $A$ and by default
PETSc uses the already assembled operator $A$ as a preconditioner. This can be
specified in code as shown in
\cref{fig:mass_solve}.

\begin{figure}
\begin{lstlisting}[language=python,linerange=65-68,firstnumber=65]
options.prefixPush("fieldsplit_0_")
options["ksp_type"] = "preonly"
options["pc_type"] = "bjacobi"
options.prefixPop()  # fieldsplit_0_
\end{lstlisting}
\caption{Continuation of \cref{fig:petsc_wrapped_code}. We tell PETSc to
	approximate $\tilde{A}^{-1}$ using one application of block Jacobi
	preconditioning with the matrix $A$, the already-assembled upper-left block of
	$K$.}\label{fig:mass_solve}
\end{figure}

\begin{figure}
\begin{lstlisting}[language=python,linerange=70-111,firstnumber=70]
n = FacetNormal(domain)
alpha = fem.Constant(domain, 4.0)
gamma = fem.Constant(domain, 8.0)
h = CellDiameter(domain)

s = -(
    inner(grad(p), grad(p_t)) * dx
    - inner(avg(grad(p_t)), jump(p, n)) * dS
    - inner(jump(p, n), avg(grad(p_t))) * dS
    + (alpha / avg(h)) * inner(jump(p, n), jump(p_t, n)) * dS
    - inner(grad(p), p_t * n) * ds
    - inner(p * n, grad(p_t)) * ds
    + (gamma / h) * p * p_t * ds
)

S = assemble_matrix(fem.form(s))
S.assemble()


class SchurInv:
    def setUp(self, pc):
        self.ksp = PETSc.KSP().create()
        self.ksp.setOptionsPrefix(pc.getOptionsPrefix() + "SchurInv_")
        self.ksp.setOperators(S)
        self.ksp.setFromOptions()

    def apply(self, pc, x, y):
        self.ksp.solve(x, y)


options.prefixPush("fieldsplit_1_")
options["ksp_type"] = "preonly"
options["pc_type"] = "python"
options["pc_python_type"] = __name__ + ".SchurInv"
options.prefixPush("SchurInv_")
options["ksp_type"] = "preonly"
options["pc_type"] = "hypre"
options.prefixPop()  # SchurInv_
options.prefixPop()  # fieldsplit_1_

options.prefixPop()  # wrapped_
options.prefixPop()  # mp_
\end{lstlisting}
\caption{Continuation of \cref{fig:mass_solve}. We first setup a class
	\texttt{SchurInv} with a method \texttt{apply} that will apply the approximate
	inverse of the discontinuous Galerkin Laplacian matrix \texttt{S} to the vector
	\texttt{x} and place the result in \texttt{y}. We then tell PETSc to use this
	method when it needs the action of $\tilde{S}^{-1}$.}\label{fig:schur_solve}
\end{figure}

\begin{figure}
\begin{lstlisting}[language=python,linerange=113-116,firstnumber=113]
solver.setFromOptions()

x = fem.petsc.create_vector(L_dolfinx, kind="mpi")
solver.solve(b, x)
\end{lstlisting}
\caption{Continuation of \cref{fig:schur_solve}. Finally, we set all of the
	options on the PETSc objects and solve. This solver setup gives a nearly mesh
	independent number of GMRES iterations (17) tested up to a mesh size of $1024
	\times 1024$.}\label{fig:do_the_solve}
\end{figure}

For $\tilde{S}^{-1}$ we take a single application of algebraic multigrid
preconditioned discontinuous Galerkin approximation of the Laplacian on the
finite element pressure space $P_h$. This can be specified as in
\cref{fig:schur_solve}. This choice is justified by the spectral equivalence
between the Schur complement and the Laplacian, for further details see
e.g.~\cite{benzi_numerical_2005}.
In the final step, we complete the solver setup and solve the problem using the sequence
of commands shown in \cref{fig:do_the_solve}.

We remark that the same configuration of the linear solver can be achieved
using the alternative DOLFINx's assembly routines that produce the PETSc
block-structured matrix of type MATNEST~\cite{brown_composable_2012}, and without
the use of the utilities provided by FEniCSx-pctools. The main advantage
offered by FEniCSx-pctools is the possibility to change the preconditioner
setup at runtime without the need to modify the model specification (typically,
the arrangement of finite element function spaces that determines the block
layout of the system matrix).

\section*{Use in research}

In this section we show scaling results demonstrating the relevance of the
software to solving real-world problems involving the solution of partial
differential equations (PDE) at scale on high-performance computers (HPC).

\begin{figure}[!h]
\centering
\vspace{2mm}
\includegraphics[width=12cm]{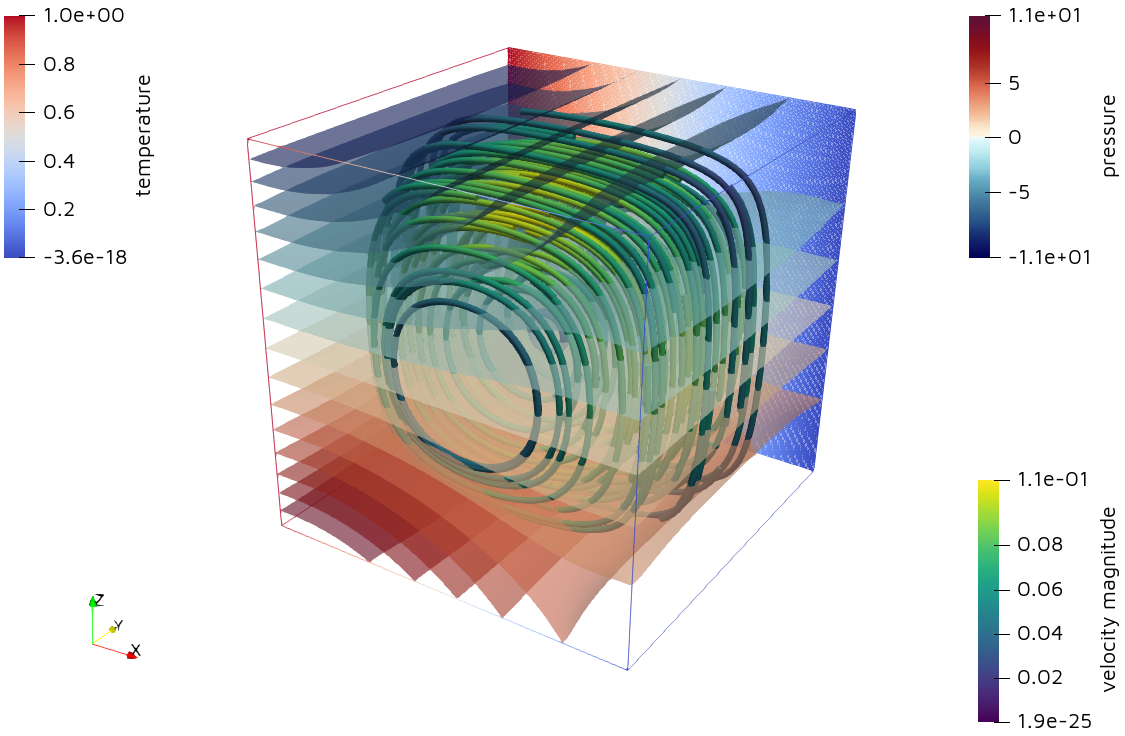}
\caption{Visualisation of the solution to the Rayleigh-B\'{e}nard
	problem~\cite{kirby_solver_2018}. The fluid velocity
	is shown as streamlines coloured by the velocity magnitude, the fluid
	pressure as an isocontour field, and the fluid temperature on the back
	surface only. Fluid is held at a higher temperature on the left and a
	lower temperature on the right of the cube. Less dense fluid rises on
	the left, and denser fluid sinks on the right (the setup takes
	the gravitational acceleration vector pointing up). At steady-state this
	forms a distinctive circulation pattern which is clearly
	shown in the fluid velocity streamlines.}
\label{fig:rayleigh-benard}
\end{figure}

We perform weak scalability tests up to \num{8192} MPI (Message Passing Interface)
processes on a three-field (temperature, velocity and pressure) problem, describing
a steady-state incompressible flow with thermal convection, preconditioned with
algebraic multigrid for the temperature block and PCD approach for the velocity-pressure
Navier-Stokes blocks~\cite{elman_finite_2014}. The PDE and problem data (domain, boundary
conditions etc.) are exactly the same as those described
as the Rayleigh-B\'{e}nard problem in~\cite{kirby_solver_2018} and solved with
Firedrake~\cite{rathgeber_firedrake_2016}, so for brevity's sake we do not
repeat the details. The solution is described and visualised in
\cref{fig:rayleigh-benard}. Compared with~\cite{kirby_solver_2018} we use a
slightly different design for the PCD component of our preconditioner
following~\cite{blechta_towards_2019}. This results in fewer outer
Newton-Raphson iterations and fewer inner Krylov solver iterations than the
design proposed in~\cite{kirby_solver_2018}.

The weak scalability tests were performed on the University of Luxembourg
Aion cluster~\cite{varrette_management_2022}. The Aion cluster consists of 354
nodes each with two 64 core AMD Epyc ROME 7H12 processors attached to 256 GB
RAM. The nodes are connected with an Infiniband HDR network in a `fat-tree'
topology. We built DOLFINx 0.7.0 against PETSc 3.20.0, OpenMPI 4.0.5 and Python
3.8.6 using the GCC 10.2.0 compiler suite with \texttt{-march=znver2 -O3}
optimisation flags. All experiments were performed at 50\% core utilisation per
node, i.e. 64 cores per node, as these low-order finite element problems are
typically memory-bandwidth constrained. The node allocation was taken in
exclusive mode, i.e. with no other jobs running on the same node.

In the weak scaling test we solve the problem on 1 node and double the number
of nodes until we reach 128 nodes (\num{8192} MPI processes). We simultaneously
increase the problem size so that the number of degrees of freedom (DOF) per
MPI rank remains fixed at around \num{100000}. A breakdown of performance
metrics for the weak scaling study is shown in \cref{table:weak_scaling}. The
last column shows the total wall time to execute the
\texttt{SNES.solve} method, which includes the DOLFINx assembly of the
necessary linear operators and their PETSc preconditioning and solution. In
summary, the time for solution stays roughly constant,
demonstrating the excellent combined weak scaling of DOLFINx, PETSc and the
coupling implemented by FEniCSx-pctools. The code to execute these experiments
and our raw timing data is available in the supplementary material in the
\texttt{examples/rayleigh-benard-convection} directory.

\begin{table}[!h]
\centering
\vspace{2mm}
\begin{footnotesize}
\begin{tabular}{c|c|c|c|c|c|c}
DOF & MPI & Nonlinear & Linear & Navier-Stokes & Temperature & Wall time for\\
($\times 10^6$) & processes & iterations & iterations & iterations & iterations & SNES solve (s)\\
\hline
6.359 & 64 & 2 & 8 & 115 (14.4) & 49 (6.1) & 26.7\\
12.6 & 128 & 2 & 8 & 117 (14.6) & 49 (6.1) & 27.2\\
25.64 & 256 & 2 & 9 & 133 (14.8) & 56 (6.2) & 31.2\\
101.7 & 1024 & 2 & 7 & 103 (14.7) & 43 (6.1) & 25.9\\
203.5 & 2048 & 2 & 7 & 102 (14.6) & 44 (6.3) & 26.1\\
408.9 & 4096 & 2 & 5 & 82 (16.4) & 31 (6.2) & 22.4\\
816.8 & 8192 & 2 & 6 & 102 (17) & 41 (6.8) & 27.1
\end{tabular}
\end{footnotesize}
\caption{Performance metrics for the Rayleigh-Bénard problem~\cite{kirby_solver_2018} with
	customised PCD-AMG preconditioning. Weak scaling at 100k (DOF) per process. Aion Cluster,
	50\% utilisation. The number in brackets is the average iterations per
	outer linear solve. Wall time for SNES solve is the time taken to execute
	\texttt{SNES.solve}, which includes PETSc linear algebra operations,
	DOLFINx assembly operations and the coupling implemented by FEniCSx-pctools.}\label{table:weak_scaling}
\end{table}

\section*{Quality control}

FEniCSx-pctools contains unit tests that assert that the package functions
correctly. In addition, there are three demo problems with checks for
correctness. These tests are run as part of a continuous integration pipeline.
Users can run these tests themselves by following the instructions in the
\texttt{README.rst} file. The package is fully documented and Python
type-hinted/checked.

\section*{Availability}

\subsection*{Operating system}

DOLFINx, and consequently FEniCSx-pctools, can be built on any modern
POSIX-like system, e.g. macOS, Linux, FreeBSD etc. DOLFINx can be built
natively on Windows, but at the present time only without PETSc, so
FEniCSx-pctools cannot be used on Windows.

\subsection*{Programming language}

FEniCSx-pctools is written in Python and is compatible with the CPython
intepreter version 3.10 and above.

\subsection*{Additional system requirements}

FEniCSx-pctools and its main dependencies DOLFINx and PETSc are designed with
scalability on parallel distributed memory systems using MPI. Consequently, they
can run on laptops through to large HPC systems.

\subsection*{Dependencies}

FEniCSx-pctools depends on the Python interface to DOLFINx compiled with
PETSc/ petsc4py support. Because of the many ways to install DOLFINx and
PETSc/petsc4py we point users to the upstream DOLFINx instructions.
Dependencies and optional dependencies are specified in the standard Python
packaging configuration file \texttt{pyproject.toml}. We aim to make tagged
releases of FEniCSx-pctools that are compatible with DOLFINx releases.

\section*{Reuse potential}
The design of parameter and discretisation robust block preconditioning
strategies is an active research topic in numerical analysis and computational
sciences. We can point to recent developments in designs for the
Navier-Stokes~\cite{farrell_reynolds-robust_2021} equations, poroelasticity
equations~\cite{chen_robust_2020}, magnetohydrodynamic
equations~\cite{laakmann_augmented_2022} and multiphysics interface
problems~\cite{budisa_rational_2023}. Together, FEniCSx-pctools, DOLFINx and
PETSc support the straightforward expression and testing of these
preconditioning strategies in code, and therefore are useful for researchers
who wish to quickly verify the performance of their preconditioning designs. In
addition, block preconditioning strategies are an important tool for solving
large real-world problems in computational sciences and engineering.

\section*{Supplementary material}

{\bf Archive}
\begin{itemize}
\item Name: Figshare \cite{rehor2022}
\item Persistent identifier: \url{https://doi.org/10.6084/m9.figshare.21408294.v6}
\item Licence: LGPLv3 or later
\item Publisher: Martin \v{R}eho\v{r} on behalf of Rafinex S.\`{a} r.l.
\item Version published: Version 6 created from git tag \texttt{v0.10.0.dev0-paper}
\item Date published: 04/09/2025 (ongoing)
\end{itemize}

\noindent{\bf Code repository}
\begin{itemize}
\item Name: gitlab.com
\item Persistent identifier: \url{https://gitlab.com/rafinex-external-rifle/fenicsx-pctools}
\item Licence: LGPLv3 or later
\item Date published: 27/10/2022 (ongoing)
\end{itemize}

\section*{Acknowledgements}

We would like to thank both reviewers and the editor for their thorough
review of this work which led to a significantly improved final article.

The experiments presented in this work were carried out using the HPC
facilities of the University of Luxembourg~\cite{varrette_management_2022} (see
\url{https://hpc.uni.lu}).

Jack S. Hale has a family member that works at Rafinex S.\`{a} r.l. This family
member was not involved in this research project. Martin \v{R}eho\v{r} declares no
competing interests.

\section*{Funding statement}

This research was funded in whole, or in part, by the Luxembourg National
Research Fund (FNR), grant reference RIFLE/13754363. For the purpose of open
access, and in fulfilment of the obligations arising from the grant agreement,
the author has applied a Creative Commons Attribution 4.0 International (CC BY
4.0) license to any Author Accepted Manuscript version arising from this
submission.

\printbibliography

\end{document}